
\input phyzzx
\input tables
\def\del {\partial}
\def\grad{\nabla}

\def\IGPP{\address{Institute for Geophysics and Planetary Physics,
      Lawrence Livermore National Laboratory, Livermore, CA 94550}}
\def\CFPA{\address{Center for Particle Astrophysics, University
      of California, Berkeley, CA 94720}}

\PHYSREV
\singlespace

\pubnum{CfPA-TH-92-21}
\pubtype{}
\date{July 9, 1992}
\titlepage
\singlespace
\title{\bf COBE's Constraints on the Global Monopole and Texture Theories
       of Cosmic Structure Formation}
\author{David P. Bennett}
\IGPP
\andauthor{Sun Hong Rhie}
\CFPA

\abstract
\singlespace
We report on a calculation of large scale anisotropy in the cosmic microwave
background radiation in the global monopole and texture models for cosmic
structure formation. We have evolved the six component linear gravitational
field along with the monopole or texture scalar fields numerically in an
expanding universe and performed the Sachs-Wolfe integrals directly
on the calculated gravitational fields.  On scales $> 7^\circ$,
we find a Gaussian distribution with an approximately scale invariant
fluctuation spectrum. The $\Delta T/T$ amplitude is a factor of 4-5 larger
than the prediction of the standard CDM model with the same Hubble
constant and density fluctuation normalization. The recently
reported COBE-DMR results imply that global monopole and texture models
require high bias factors or a large Hubble constant in contrast to standard
CDM which requires very low $H_0$ and bias values.  For
$H_0 = 70 {\rm {km\over sec} Mpc^{-1}}$, we find that normalizing to
the COBE results implies $b_8 \simeq 3.2\pm 1.4$ (95\% c.l.).
If we restrict ourselves
to the range of bias factors thought to be reasonable before the announcement
of the COBE results, $1.5 \lsim b_8 \lsim 2.5$, then
it is fair to conclude that global monopoles and textures are consistent
with the COBE results and are a {\it better} fit than Standard CDM.

\submit{The Astrophysical Journal Letters}
\endpage

\FIG\Monmap{A $\Delta T/T$ map for the global monopole model generated by our
    global field evolution code is displayed in a full-sky
    equal area projection. The scale is given in units of $Gv_m^2$.}
\FIG\Texmap{A $\Delta T/T$ map for the texture model generated by our
    global field evolution code is displayed in a full-sky
    equal area projection. The scale is given in units of $Gv_t^2$.}
\FIG\Powspec{The average $\Delta T/T$ power spectrum is plotted for 12 $100^3$
    monopole simulations and 12 $100^3$ texture simulations. The solid and
    dashed curves give the best fit to the power spectrum derived for
    a Harrison-Zel'dovich spectrum of primordial adiabatic density
    perturbations. The error bars give the RMS deviation from the mean,
    so they reflect the expected deviation for a single realization. }
\FIG\MonHisto{A histogram of the pixels in convolved $\Delta T/T$ map shown
    in Fig. \Monmap. The bins on the edges of the histogram include
    all the points outside the limits of the figure. The dashed curve
    is the histogram for a Gaussian with the same RMS.}
\FIG\TexHisto{A histogram of the pixels in convolved $\Delta T/T$ map shown
    in Fig. \Texmap. The bins on the edges of the histogram include
    all the points outside the limits of the figure. The dashed curve
    is the histogram for a Gaussian with the same RMS.}

\unnumberedchapters
\chapter{}

Topological defects which formed in a phase transition in the
early universe provide an attractive mechanism for the generation of
density perturbations which can grow to form galaxies and large scale
structure. Cosmic strings (Vilenkin, 1980, Zel'dovich, 1980),
global monopoles (Bariola and Vilenkin, 1989, Bennett and Rhie, 1990), and
global textures (Turok, 1989) have all been proposed as
possible seeds for cosmic structure formation.
These theories are characterized by a single adjustable parameter,
the Grand Unified theory symmetry breaking scale, $v$, and the value
($v \sim 10^{16}$ GeV, $Gv^2\sim 10^{-6}$) predicted by particle physics
(Amaldi \etal, 1991) also gives the correct amplitude to
generate galaxies and large scale structure. In contrast, inflationary models
generally cannot produce a reasonable perturbation amplitude without
a rather extreme fine tuning of the coupling constant
($\lambda \sim 10^{-12}$).

Recent studies
(Cen \etal, 1991, Park \etal, 1991, Gooding \etal, 1991 and 1992, and Turok and
Spergel, 1990) of texture seeded density perturbations in a universe
dominated by cold dark matter (CDM) have indicated that this theory
may solve two of the perceived problems with the standard CDM model:
the lack of sufficient large scale structure and quasars at high redshift.
Their results generally agree with our
calculations of global monopole (GM) and texture (T) seeded
structure formation (Bennett, Rhie, and Weinberg, 1992),
but we also find that these nongaussian seeds tend to generate large
galaxy pair velocities and cluster
velocity dispersions (see also Bartlett, Gooding and
Spergel, 1992). This can
be alleviated by selecting a larger ``bias" factor (\ie,  a lower
normalization of the density field).

Another major difference between the nongaussian
GM \& T models and the gaussian inflationary models can be seen in the
cosmic microwave background radiation (CMB) anisotropies.
In inflation scenarios, $\Delta T/T$ is due to
remnant quantum fluctuations crossing the horizon at last scattering,
and analytic results for $\Delta T/T$ have been obtained for both the scalar
mode responsible for the growth of cosmic structure
(Bond and Efstathiou, 1987)
and the tensor modes (gravity waves) (Abbott and Wise, 1984). For the
standard exponential inflation models, only the scalar growing mode
is important.
With topological defects, metric fluctuations are  generated
by the relativistic dynamics of the defects inside the horizon, and
$\Delta T/T$ is affected by the fluctuations in all six independent
components of the gravitational field. Thus, we expect the topological
seed models to predict larger $\Delta T/T$ for a fixed amplitude of
density fluctuations. The recent detection of CMB anisotropy
by the COBE DMR experiment (Smoot, \etal, 1992)
at a level somewhat higher than the prediction of the Standard CDM model
should be regarded as encouraging for topological seed models.
(Some non-standard inflationary models may also have significant tensor mode
perturbations (Davis, \etal, 1992), but they are certainly not the
{\it only} theory with tensor modes.)

A previous estimation of $\Delta T/T$
for the texture model (Turok and Spergel, 1990) was based
on a simple analytic model of a single texture evolution. They found
a non-Gaussian distribution of hot and cold spots at a level that
seems to conflict with the COBE data.
In this paper, we present realistic numerical calculations of $\Delta T/T$
on COBE scales in the GM\&T models with no simplifying assumptions.

$\Delta T/T$ on COBE scales reflects the variations in time
delay (frequency shift) along the photon paths from last scattering until
the present. This is the generalized
Sachs-Wolfe effect (Sachs and Wolfe, 1967) where not only scalar growing
mode but all gravitational field components contribute to the temperature
fluctuations.
If we choose a coordinate system such that the metric is
$g_{\mu\nu}(x)=a(\eta)^2[\eta_{\mu\nu}+h_{\mu\nu}(x)]$,
where $\eta_{\mu\nu}= {\rm diag}(-,+,+,+)$ and $h_{\mu\nu}$ is the
metric perturbation, and choose a gauge $h_{0\nu} = 0$
(Veeraraghavan and Stebbins, 1990), then
the temperature fluctuation is given by
  $$ {\Delta T\over T} = - {1\over 2} \int d\eta \hat x_i \hat x_j
           {\del h_{ij} \over \del \eta} \ , \eqn\dtstint $$
where $\hat x_i$ is the normal vector along the line of sight.
Because GM\&T  predict the early formation of objects such as quasars,
we assume the universe was reionized at high redshift and
take this into account by introducing a visibility function,
$$ f(z) = e^{h\Omega_b(1-(1+z)^{3/2})/21.7} \ ,
\eqn\visfunc $$
which measures the fraction of photons present at redshift $z$ that will
reach $z=0$ without undergoing Compton scattering. If we assume that
the electrons which scatter each photon see $\VEV{\Delta T/T} = 0$, then
we can account for reionization by inserting $f(z)$ inside the ``Sachs-Wolfe"
integral, eq. \dtstint. This means that our $\Delta T/T$ results will now
depend on $h=H_0/100 {\rm {km\over sec} Mpc^{-1}}$
and the baryon density $\Omega_b$.
We have done most of our calculations for $h\Omega_b = 0.1$ which is
about the largest plausible value. For $h\Omega_b=0.04$,
$\Delta T/T$ is only 3\% larger on COBE scales.

We evolve the source fields according to the field equations of motion
in the Freedman-Robertson-Walker background and calculate the metric
perturbations due to the energy-momentum of the scalar field
by solving the linearized Einstein equations.
The field equations are
$$ \ddot\phi^p + 2{\dot a\over a}\dot\phi^p - \grad^2\phi^p
   + a^2 \lambda\left(\phi^2-v^2\right)\phi^p = 0 \ ,
\eqn\Leq $$
where $v$ is the vacuum expectation value of the field, and
$p$ runs from 1-3 for monopoles and 1-4 for textures.
Because the defect core size is very much smaller than the our grid
spacing, we can use the equation for the $\lambda\rightarrow
\infty$ and improve the dynamic range of the calculations
(Bennett and Rhie, 1990),
$$ \left(\delta_{pq} - {\phi^p\phi^q\over v^2}\right)
   \left(\ddot\phi^q + 2{\dot a\over a}\dot\phi^q - \grad^2\phi^q\right)
   = 0 \ .
\eqn\NLeq $$

The energy-momentum tensor for the scalar fields is given by
$$ \eqalign{\Theta_{00} =& {1\over 2}\dot{\vec\phi}^2
            + {1\over 2}\left(\grad\vec\phi\right)^2 \ , \cr
 \Theta_{0i} =& \dot{\vec\phi}\cdot\del_i\vec\phi \ , \cr
 \Theta_{ij} =& \del_i\vec\phi\cdot\del_j\vec\phi + {1\over 2}
   \delta_{ij}\left( {1\over 2}\dot{\vec\phi}^2
            - {1\over 2}\left(\grad\vec\phi\right)^2 \right) \ , \cr }
\eqn\Thetas $$
and the linearized Einstein equations are (Veeraraghavan and Stebbins, 1990)
$$ \ddot h + {\dot a\over a}\dot h + 3 {\dot a^2\over a^2} \delta_c
   = -8\pi \left( \Theta_{00} + \Theta \right) \ ,
\eqn\heqn $$
$$\eqalign{ \ddot{\tilde h}_{ij} - \grad^2\tilde h_{ij} + 2 {\dot a\over a}
    \dot{\tilde h}_{ij} - {1\over 3}\del_i\del_j h
    + {1\over 9}\delta_{ij}\grad^2 h& \cr
    + \del_i\del_k\tilde h_{jk} + \del_j\del_k\tilde h_{ik}
    - {2\over 3} \delta_{ij} \del_k\del_l \tilde h_{kl}
    &= 16\pi\tilde\Theta_{ij} \ , \cr  }
\eqn\hijeqn $$
where $h$ and $\Theta$ refer to the traces of $h_{ij}$ and $\Theta_{ij}$.
The $\tilde {}$ is used to refer to the trace-free terms
$\tilde h_{ij} = h_{ij} - {1\over 3}\delta_{ij} h$ and
$\tilde \Theta_{ij} = \Theta_{ij} - {1\over 3}\delta_{ij} \Theta$.
The perturbation in cold dark matter $\delta_c$ obeys
$$ \dot{\delta}_c = -{1\over 2} \dot h \ ,
\eqn\dceqn $$
and the following (non-dynamical) constraint equations must be satisfied,
$$ \eqalign{ &{1\over 2} \del_i\del_j \tilde h_{ij} + {1\over 3}\grad^2 h
   = 8\pi\Theta_{00}
   + 3\left({\dot a\over a}\right)^2 \delta_c - {\dot a\over a} \dot h \ , \cr
&{1\over 2} \del_j \dot{\tilde h}_{ij} - {1\over 3}\del_i \dot h
     = 8\pi\Theta_{0i} \ . \cr }
\eqn\constraint $$

In theories with ``external" sources such as global monopoles or textures,
these equations impose important constraints on the initial conditions.
In particular, since $\Theta_{00}$ cannot vanish if we are to have interesting
density perturbations, the initial energy density fluctuations in the
$\vec\phi$ field must be ``compensated" by fluctuations of the opposite sign
in the other fields. For the calculations reported in this paper, we have
taken $h(\eta_0)=0$, $\dot{h}(\eta_0)=0$, $\tilde h_{ij}(\eta_0)=0$,
$\dot{\tilde h}_{ij}(\eta_0)=0$, $\dot{\vec\phi}(\eta_0)=0$, and
$\delta_c(\eta_0) = -(8\pi/3)(a/\dot{a})^2 \Theta_{00}(\eta_0)$. Once the
constraint equations, \constraint, are satisfied initially, they will be
satisfied at subsequent times if the equations of motion, \heqn-\dceqn,
are satisfied. Extreme care must be taken when evolving these equations
numerically, because small numerical errors can excite growing mode solutions
on scales outside the horizon where the physical modes do not grow. When these
scales finally come inside the horizon, the errors can have grown large enough
to compete with the physical perturbations that are generated by the source
inside the horizon. We have found that we can keep these errors under control
even when $\vec\phi$ takes random values on the scale of 1 grid spacing
by going to extremely small time-steps ($\Delta t \approx 0.01 \Delta x$).
In order to satisfy the second equation in \constraint, we find that global
monopole and texture fields must be sufficiently smooth, and this limits us
to an initial horizon size of $\gsim 8$ grid spacings.

Another difficulty with evolving eqs. \heqn-\dceqn\ numerically is that
the mixed partial derivatives in \hijeqn\ make it difficult to come up with
an {\it explicit} differencing scheme that is stable. {\it Implicit}
differencing schemes would be prohibitively expensive in computer time
because we would still have to take very small time-steps to avoid the
super-horizon scale growing mode solution discussed above since the
growing mode is a physical mode, {\it not} a purely numerical one.

Fortunately, we have found it possible to remove the offensive mixed partial
into \hijeqn. This substitution yields
$$ \eqalign{ \ddot{\tilde h}_{ij} =& \grad^2\tilde h_{ij}
    - 2{\dot a\over a}\dot{\tilde h}_{ij} - \del_i\del_j h
    + 16\pi\del_i \vec\phi \cdot \del_j \vec\phi \bigg|_{\eta=\eta_0}
    - 32\pi\int^\eta_{\eta_0} d\eta\,\dot{\vec\phi}\cdot\del_i\del_j\vec\phi
                                      \cr
    +& \delta_{ij}\left[ {1\over 3}\grad^2 h + {16\pi\over 3} \dot{\vec\phi}^2
     + 4\left({\dot a\over a}\right)^2 \delta_c
     - {4\over 3}{\dot a\over a}\dot h \right] \ , \cr }
\eqn\hijeqnnew $$
for eq. \hijeqn. The mixed partial derivatives of $h$ and $\vec\phi$ in eq.
\hijeqnnew\ do not give rise to numerical instabilities because the equations
of motion for $h$ and $\vec\phi$
(\heqn\ and \NLeq) do not contain any mixed partial derivatives.

Our numerical simulations use a modified second order leapfrog scheme to
evolve eqns. \NLeq, \heqn, \dceqn, and \hijeqnnew\ in time. $\Delta T/T$
is determined by integrating eq. \dtstint\ (with the visibility function
\visfunc\ inserted) along photon trajectories that converge to a point
at the end of the simulation. We have done about 25 simulations on $64^3$
grids and 12 simulations on $100^3$ grids for each of the global monopole
and texture models. The RMS $\Delta T/T$ values for the $100^3$ grids are
$17.80\pm 2.00 Gv^2_m$ and $10.29\pm 1.43 Gv^2_t$ for monopoles and
textures respectively. ($v_m$ and $v_t$ refer to the vacuum expectation
values of the monopole and texture fields.) These numbers can be compared
to the RMS fluctuation measured by COBE
$(\Delta T/T)_{RMS} = 1.10\pm 0.18 \times 10^{-5}$
to yield $Gv_m^2 = 6.18\pm 1.23 \times 10^{-7}$ for global monopoles or
$Gv_t^2 = 1.07\pm 0.23 \times 10^{-6}$ for textures if we assume that
monopoles or textures are responsible for the $\Delta T/T$ observed by COBE.

Figs. \Monmap\ and \Texmap\ show the simulated
full-sky temperature maps for the global monopole and texture models
respectively smoothed to the same $10^\circ$ scale as the COBE maps.
The scale on these plots is given in terms of $Gv^2$.
Figs. \MonHisto\ and \TexHisto\ show histograms of the $\Delta T/T$
distributions in the monopole and texture cases. Note that after the smoothing
the maps contain only about 400 independent pixels. Thus, the departures
from a Gaussian distribution are not significant.

Fig. \Powspec\ shows the angular power spectrum, $\Delta T^2_l$, of
our simulated $\Delta T/T$ maps.
$$ {\Delta T\over T}(\theta,\phi) = \sum_{l,m} a_{lm} Y_{lm}(\theta,\phi) \ ,
\eqn\ylm $$
$$ \Delta T^2_l = {1\over 4\pi} \sum_m |a_{lm}|^2 \ .
\eqn\Ql $$
The solid and dashed curves give the
best fit to the predicted power spectrum for Harrison-Zel'dovich ($n=1$)
primordial adiabatic density perturbations (Bond and Efstathiou, 1987),
$$\Delta T_l^2 = (Q_{rms-PS})^2{(2l+1)\over 5}{\Gamma(l+(n-1)/2)\Gamma((9-n)/2)
                  \over  \Gamma(l+(5-n)/2)\Gamma((3+n)/2)} \ .
\eqn\HZpow $$
 The
best fit of the form, \HZpow, gives $n=1.1\pm 0.3$ for global monopoles
and $n=1.2\pm 0.2$ for textures when we remove the quadrupole from the
fit as was done for the COBE data. (The error bars in these fits reflect
mainly systematic errors.) The Harrison-Zel'dovich value, $n=1$, gives a
good fit to our simulations and to the COBE data, so it makes sense to
compare the fit amplitudes. The COBE value is
$Q_{rms-PS} = 6.11\pm 1.46\times 10^{-6}$, and we obtain
$Q_{rms-PS} = 8.7\pm 1.6 Gv_m^2$ for monopoles and
$Q_{rms-PS} = 4.7\pm 0.5 Gv_t^2$ for textures. A comparison of these values
gives $Gv_m^2 = 7.0\pm 2.1 \times 10^{-7}$ and
$Gv_t^2 = 1.30\pm 0.34 \times 10^{-6}$ consistent with the values obtained
above.

It is worth noting that topological defect theories generically predict
$\Delta T/T$ and $\delta\rho/\rho$ spectra that are slightly steeper
than Harrison-Zel'dovich on very large scales ($\gsim 1000$ Mpc).
The reason for this
is that with topological defects, unlike inflation, the gravitational
fields are generated inside the horizon, so that scales near the horizon have
yet to receive their full ``share" of perturbations.
Thus, the largest scales are expected to have less power
than the scale-free spectrum (\ie\ $n>1$). (This effect is
partially compensated for by the effects of reionization which reduce
$\Delta T/T$ on small scales.) For cosmic strings,
the effect should be even more pronounced since the coherence scale of
the strings is smaller than that of global monopoles or textures
(Bennett, Stebbins, and Bouchet, 1992). In contrast, power law
inflationary models (Davis, \etal, 1992)
which are  able to fit the COBE results with a
reasonable value for $b_8$ predict $n < 1$. Thus, if the four year COBE
results converge to $n=1.50\pm 0.25$, it will be fair to conclude that
topological defect models are preferred over quantum fluctuations during
inflation as the source of the primordial density perturbations.
\nextline

\begintable
   | \multispan{2}\tstrut\hfil\ Monopoles \hfil |
     \multispan{2}\tstrut\hfil\ Textures  \hfil\cr
 $h$ | $Gv_m^2$ | $b_8$ | $Gv_t^2$ | $b_8$ \crthick
$0.5$ | $2.49 \times 10^{-6}/b_8$ | $4.03\pm 0.80$
      | $4.56 \times 10^{-6}/b_8$ | $4.27\pm 0.98$ \nr
$0.6$ | $2.19 \times 10^{-6}/b_8$ | $3.54\pm 0.70$
      | $3.93 \times 10^{-6}/b_8$ | $3.68\pm 0.84$ \nr
$0.7$ | $1.96 \times 10^{-6}/b_8$ | $3.17\pm 0.63$
      | $3.48 \times 10^{-6}/b_8$ | $3.25\pm 0.75$ \nr
$0.8$ | $1.82 \times 10^{-6}/b_8$ | $2.94\pm 0.58$
      | $3.17 \times 10^{-6}/b_8$ | $2.97\pm 0.68$ \nr
$1.0$ | $1.64 \times 10^{-6}/b_8$ | $2.65\pm 0.53$
      | $2.75 \times 10^{-6}/b_8$ | $2.57\pm 0.59$
    \endtable
\noindent{\it Table 1.} The scalar field normalizations and best fit
bias parameters, $b_8$, to the COBE-DMR RMS anisotropy at $10^\circ$
are tabulated as a function of $h$. 1 $\sigma$ errors are reported.

In order to translate our results into limits on theories of cosmic
structure formation, we must normalize
$Gv^2$ to give a reasonable spectrum of density perturbations. Because of
uncertainties in the relationship between the number density of galaxies
and the mass density, this normalization is conventionally given in terms
of a bias factor, $b_8$, such that the RMS mass density fluctuation $= 1/b_8$
after smoothing with an $8h^{-1}$Mpc radius top hat. Table 1 gives
these normalizations as determined in Bennett, Rhie, and Weinberg (1992).

Table 1 also lists the predicted $b_8$ values as a function of
$h = H_0/100{\rm km\over sec}{\rm Mpc^{-1}}$ as determined by a comparisons
of the predicted RMS fluctuation at $10^\circ$.
We can see that the predicted $b_8$ values run very high for small
values of the Hubble constant. Choosing a large Hubble constant in
the $\Omega = 1$ universe that we have assumed is problematic due to
the implied very short age for the universe, but with $h = 0.7$
(the smallest value that is consistent with most of the measured values of $h$)
$b_8=2.5$ is within $\sim 1 \sigma$ of the mean in the both the global
monopole and texture models. Thus, global monopoles and textures with
a bias factor in the range of $2.5-3$ seem to be in reasonably good agreement
with the COBE data. In a separate study of large
scale structure in the global monopole and texture models,
(Bennett, Rhie, and Weinberg, 1992), we find that these high bias global
monopole and texture models do reasonably
well in matching the observed large scale structure. Further work is required
to see if the required biasing can be obtained dynamically, however.

Finally, let us compare our results to those of other, well motivated
theories of large scale structure formation. We find that for similar
values of $h$ and $b_8$, global monopole and textures predict $\Delta T/T$
on COBE scales that is a factor of 4-5 larger than the standard CDM
prediction (Bond and Efstathiou, 1987). With a reasonable bias value,
$1.5 \gsim b_8 \gsim 2.5$, this model is inconsistent with the COBE measurement
for $h > 0.5$ and but perhaps barely consistent for $h=0.5$. If we demand that
$1.5 \gsim b_8 \gsim 2.5$ for global monopoles and textures, we find
consistency with COBE for the entire range, $0.5 < h < 1.0$ at the
$2 \sigma$ confidence level and $0.7 < h < 1.0$ at the $1\sigma$ level.
The power law inflationary models discussed by Davis, \etal, (1992)
can be made consistent with reasonable $b_8$ values because they have
significant contributions to $\Delta T/T$ from tensor
modes that do not contribute to $\delta\rho/\rho$. These models do seem to
have some difficulty with forming galaxies early enough, however (Adams, \etal,
1992). Other models in with a smaller amount of CDM, such as hot $+$ cold DM
models or low $\Omega$ models seem to fit the COBE results reasonably well
(Wright, \etal, 1992),
but they are less attractive theoretically. Cosmic Strings + HDM seem to
be a good fit to the COBE results (Bennett, Stebbins, and Bouchet, 1992),
but the theoretical error bars are presently rather large. Thus, none of
the best motivated models are singled out by the COBE results, but global
monopoles and textures are arguably the best fit to COBE among the $\Omega =1$
pure CDM models. Further work into the details of galaxy formation and
$\Delta T/T$ on smaller angular scales is certainly warranted.

\singlespace
\ack
We would like to thank A. Stebbins and D. Weinberg for helpful discussions.
This work was supported in part
the U.S. Department of Energy at the Lawrence Livermore
National Laboratory under contract No. W-7405-Eng-48
and by the NSF grant No. PHY-9109414.


\def\pp{\parshape 2 0truecm 16.25truecm 2truecm 14.25truecm}
\def\newrefout{\par \penalty-400 \vskip\chapterskip
   \spacecheck\referenceminspace \immediate\closeout\referencewrite
   \referenceopenfalse
   \line{\fourteenrm\hfil REFERENCES\hfil}\vskip\headskip
   }
\newrefout
\parskip=0pt
\pp\par
Abbott, L., and Wise, M., 1984, Nucl. Phys. {\bf B 244}, 541.
\pp\par
Adams, F. C., Bond, J. R., Freese, K., Frieman, J. A., and Olinto, A. V.,
       1992, CITA preprint.
\pp\par
Amaldi, U. \etal, 1991, Phys. Lett. {\bf B 260}, 447.
\pp\par
Bartlett, J., Gooding, A. K., and Spergel, D. N., 1992, Berkeley preprint.
\pp\par
Bariola, M., and Vilenkin, A., 1989, Phys. Rev. Lett. {\bf 63}, 341.
\pp\par
Bennett, D. P., Stebbins, A., and Bouchet, F. R., 1992, IGPP preprint
       UCRL-JC-110803, submitted to Astrophys. J. Lett.
\pp\par
Bennett, D. P., and Rhie, S. H., 1990, Phys. Rev. Lett. {\bf 65}, 1709.
\pp\par
Bennett, D. P., Rhie, S. H., and Weinberg, D. H. 1992, in preparation.
\pp\par
Bond, J. R., and Efstathiou, 1987, Mon. Not. R. Astron. Soc. {\bf 226}, 655.
\pp\par
Cen, R. Y., Ostriker, J. P., Spergel, D. N., and Turok, N., 1991, Astrophys.
J.,
       {\bf 383}, 1.
\pp\par
Davis, R., Hodges, H., Smoot, G. F., Steinhardt, P. J., and Turner, M. S.,
      1992, preprint.
\pp\par
Gooding, A. K., Spergel, D. N., and Turok, N., 1991, Astrophys. J.,
       {\bf 372}, L5.
\pp\par
Gooding, A. K., \etal, 1992, Astrophys. J., {\bf 393}, 42.
\pp\par
Park, C., Spergel, D. N., and Turok, N., 1991, Astrophys. J., {\bf 372}, L53.
\pp\par
Sachs, K., and Wolfe, A. M., 1967, Astrophys. J., {\bf 147}, 73.
\pp\par
Smoot, G., \etal, 1992, COBE preprint.
\pp\par
Turok, N., 1989, Phys. Rev. Lett. {\bf 63}, 2625.
\pp\par
Turok, N., and Spergel, D. N., 1990, Phys. Rev. Lett. {\bf 64}, 2736.
\pp\par
Veeraraghavan, S., and Stebbins, A., 1990, Astrophys. J., {\bf 365}, 37.
\pp\par
Vilenkin, A., 1980, Phys. Rev. Lett. {\bf 46}, 1169, 1496(E).
\pp\par
Wright, E., \etal, 1992, COBE preprint.
\pp\par
Zel'dovich, Y. B., 1980, Mon. Not. R. Astron. Soc. {\bf 192}, 663.

\figout\end